\documentclass[12pt]{article}
\usepackage{amsmath}
\usepackage{mathrsfs}
\usepackage{natbib,graphicx,setspace,lscape,longtable,xr}  
\usepackage{mathrsfs,amsthm,amssymb,color}
\usepackage{natbib,epsfig,graphicx,pdfpages}   

\usepackage{rotating}
\usepackage{float}
\usepackage{bm}
\usepackage{ragged2e}
\usepackage{setspace}
\usepackage{threeparttable}
\usepackage{multirow}
\usepackage[shortlabels]{enumitem}
\usepackage{etoolbox}
\usepackage{subfig}
\usepackage[ruled]{algorithm2e}
\usepackage{adjustbox}
\usepackage{blindtext}
\usepackage{xcolor}
\usepackage{hyperref}
\hypersetup{
	linkcolor  = blue,
	citecolor  = blue,
	urlcolor   = blue,
	colorlinks = true,
}

\apptocmd{\thebibliography}{\setlength{\itemsep}{0pt}}{}{} 

\bibpunct{(}{)}{;}{a}{,}{,}

\newtheorem{theorem}{Theorem}
\newtheorem{lemma}{Lemma}
\newtheorem{proposition}{Proposition}
\newcommand{\mylabel}[2]{#2\def\@currentlabel{#2}\label{#1}}
\newcommand{\csection}[1]
{\begin{center}
		\stepcounter{section}
		{\bf\large\arabic{section}. #1}
	\end{center}
}

\newcommand{\csubsection}[1]{
	\begin{center}
		\stepcounter{subsection}
		{\it\arabic{section}.\arabic{subsection}. #1}
	\end{center}
}

\def\ve{\varepsilon}

\def\beq{\begin{equation}}
	\def\eeq{\end{equation}}
\def\beqr{\begin{eqnarray}}
	\def\eeqr{\end{eqnarray}}
\def\beqrs{\begin{eqnarray*}}
	\def\eeqrs{\end{eqnarray*}}
\def\bet{\begin{theorem}}
	\def\eet{\end{theorem}}
\def\bel{\begin{lemma}}
	\def\eel{\end{lemma}}
\def\bep{\begin{proposition}}
	\def\eep{\end{proposition}}
\def\bg{\begin{figure}[tbph]\begin{center}}
		\def\eg{\end{center}\end{figure}}

\def\bc{\begin{center}}
	\def\ec{\end{center}}

\def\wt{\widetilde}
\def\wh{\widehat}

\def\var{\mbox{var}}

\DeclareMathOperator*{\argmax}{argmax}
\DeclareMathOperator*{\argmin}{argmin}

\def\diag{\mbox{diag}}

\def\1{\mbox{\boldmath $1$}}
\def\0{\boldsymbol{0}}

\def\mL{\mathcal L}
\def\mM{\mathcal M}

\def\mP{\mathcal P}

\def\mbR{\mathbb R}
\def\mS{\mathcal S}

\def\S{\mathbb S}

\newcommand{\bbeta}  {\boldsymbol{\beta}}

\renewcommand{\epsilon}{{\ve}}
\renewcommand{\hat}{\widehat}
\def\wt{\widetilde}

\def\thetaJDS{\widehat{\theta}_{\textup{JDS}}}
\def\thetaoneshot{\wh\theta_{\operatorname{os}} }

\def\thetam{\wh \theta_{(m)}}

\def\thetap{\wh \theta_{\operatorname{stage,0}} }

\def\thetaos{\wh\theta_{\operatorname{stage,1}}}

\usepackage[a4paper,bindingoffset=0.2in,%
left=1in,right=1in,top=1in,bottom=1in,%
footskip=.25in]{geometry}

\def\boxit#1{\vbox{\hrule\hbox{\vrule\kern6pt\vbox{\kern6pt#1\kern6pt}\kern6pt\vrule}\hrule}}

\numberwithin{equation}{section}

\begin{document}

\begin{center}
	{\bf\large A Selective Review on Statistical Methods for Massive Data Computation: Distributed Computing, Subsampling, and Minibatch Techniques}\\
	Xuetong Li$^{a}$, Yuan Gao$ ^{a\ast} $, Hong Chang$^{a}$, Danyang Huang$^{b}$, Yingying Ma$^{c}$,\\ Rui Pan$^{d}$, Haobo Qi$^{e}$, Feifei Wang$^{b}$, Shuyuan Wu$^{f}$, Ke Xu$^{g}$, Jing Zhou$^{b}$, \\ Xuening Zhu$^{h}$, Yingqiu Zhu$^g$, Hansheng Wang$^{a}$    \\
{\it\small
 \begin{singlespace}
		$ ^a $Guanghua School of Management, Peking University, Beijing, China;
		$ ^b $Center for Applied Statistics and School of Statistics, Renmin University of China, Beijing, China; 
        $ ^c $School of Economics and Management, Beihang University, Beijing, China;
        $ ^d $School of Statistics and Mathematics, Central University of Finance and Economics, Beijing, China;
        $ ^e $School of Statistics, Beijing Normal University, Beijing, China;
        $ ^f $School of Statistics and Management, Shanghai University of Finance and Economics, Shanghai, China;
        $ ^g $School of Statistics, University of International Business and Economics, Beijing, China;
        $ ^h $School of Data Science and MOE-Laboratory for National Development and Intelligent Governance, Fudan University, Shanghai, China
\end{singlespace}	
  }

 \begin{singlespace}
  \begin{abstract}

  This paper presents a selective review of statistical computation methods for massive data analysis. A huge amount of statistical methods for massive data computation have been rapidly developed in the past decades. In this work, we focus on three categories of statistical computation methods: (1) distributed computing, (2) subsampling methods, and (3) minibatch gradient techniques. The first class of literature is about distributed computing and focuses on the situation, where the dataset size is too huge to be comfortably handled by one single computer. In this case, a distributed computation system with multiple computers has to be utilized. The second class of literature is about subsampling methods and concerns about the situation, where the {\color{black} sample} size of dataset is small enough to be placed on one single computer but too large to be easily processed by its memory as a whole. The last class of literature studies those minibatch gradient related optimization techniques, which have been extensively used for optimizing various deep learning models.
  
  \end{abstract}
\end{singlespace}	
  

\end{center}

\begin{footnotetext}[1]{
 Yuan Gao ({\it yuan\_gao@pku.edu.cn}) is the corresponding author.}
\end{footnotetext}

\newpage
		
\csection{INTRODUCTION}

Modern statistical analysis often involves datasets of massive size \citep{fan2020statistical}, for which effective computation methods are indispensable. On one side, the huge demand for computation methods for massive data analysis places serious challenges on the traditional statistical methods, which have been developed for datasets of regular size. On the other side, it also stimulates new research efforts, which try to conquer computation challenges by statistical wisdom. The research along this direction not only benefits the real practices with massive datasets but also inspires new statistical theory. The objective of this work is to provide a selective review about this exciting research area, which has been rapidly developed during the past many years. 
Then the objective here is not to provide a complete list for all the research works related to massive data computation. This is obviously a mission impossible. Instead, we should focus on three categories of statistical computing methods. They are, respectively, distributed computing 
methods, subsampling methods, and minibatch gradient descent methods. We try to organize the most important and relevant ones associated with those three categories in a structured way, so that follow-up researchers might benefit. Due to our limited understanding about the past literature and also the space constraint of a regular manuscript, we might miss some important references therein. If that happens, our sincere apology in advance and we should be more than happy to hear the feedback.

As the title suggests, this review is a selective review about statistical methods for massive data analysis. Then, the meaning of “massive data” needs to be precisely defined. We argue that whether a dataset is massive or not is relative to the computation resource. In the most ideal situation, if one is given a super computer with an unlimited amount of hard drive and memory (both CPU and GPU memory) together with a super powerful CPU, then no dataset can be considered as massive. In this ideal situation, any dataset of any size can be easily placed on one hard drive, loaded into the memory as a whole, and then processed in no time. Unfortunately, such an ideal situation never happens in reality. In real practice, most researchers are given only a limited amount of computation resources, which put various constraints on the computation power. The constraints could be the hard drive. If the size of the data exceeds one single computer’s hard drive capacity, then a distributed system has to be used to place the data. A natural question arises immediately: should we also compute it in a distributed way? This inspires a large amount of research for distributed computing, if a powerful distributed computation system is indeed available. With such a system, we find that it might remain to be practically appealing to distribute a large dataset on a powerful distributed system, even if the data size is not strictly larger than one single computer’s hard drive. This makes the subsequent computation more convenient. This constitutes the first part of the selective literature to be reviewed in this work.

Distributed computing is a powerful solution for problems with extremely large scale datasets. However, this seems not the most typical situation. The most typical one in real practice is an embarrassing situation, where the dataset sizes are not extremely large but large enough to cause a lot of computation challenges. 
To fix the idea, consider for example a dataset of a size (for example) 100GB. Note that this is a size substantially smaller than that of a hard drive of a modern computer (e.g., 2TB), but much larger than the size of the typical memory (e.g., 32GB). For a dataset of this embarrassing size, one straightforward solution remains to be distributed computing, as mentioned in the previous paragraph. However, this straightforward solution seems not the only best one for at least two reasons. First, to implement a distributed computing algorithm, one needs a powerful distributed computation system. Depending on its size, the distributed computer system could be very expensive, if not completely not affordable. Second, to utilize a distributed computing system, appropriate programming techniques are necessarily needed. Popularly used programming frameworks (e.g., Spark, Hadoop) need to be learned. This is unfortunately a painful learning process for most field practitioners (e.g., a medical scientist), who are not professional statistical programmers.
Therefore, it seems that there is a practical need for a handy computation method, which can deal with datasets of any size on a given hard drive and can be easily implemented on one single computer. The key challenge here is how to accomplish a massive data computation task with limited memory (both CPU and GPU memories) constraints. This leads to a huge body of literature about subsampling, and/or streaming data analysis. This constitutes the second part of the literature to be reviewed this work. 

Both the problems of distributed computing and computing with memory constraint concern about the computation problems of datasets with massive sizes. 
However, we often encounter situations where not only the dataset sizes are large, but also the model sizes are extremely large.
The most typical example in this regard is various deep learning methods.
To fix the idea, consider for example the famous \textit{ImageNet} dataset of  \cite{deng2009imagenet}, which contains a total of over 1.3 million color images belonging to 1,000 classes. The total amount of images is about 150GB in size. 
Next, consider for example a classical deep learning model VGG16 \citep{simonyan2014very}. This is a convolution neural network (CNN) model with a total of over 130 million parameters.
To train this VGG16 model on the \textit{ImageNet} dataset, all the model parameters need to be fully placed in the GPU memory for fast tensor computation. 
Unfortunately, once this sophisticated VGG16 model is fully loaded into the GPU memory, the space left for data processing is inevitably significantly reduced. 
Consequently, the \textit{ImageNet} dataset has to be processed in a minibatch-by-minibatch manner. Here, a minibatch refers to a small or even tiny subset of the whole sample. 
Depending on the way this subsample is generated, we might have streaming data based minibatches \citep{chen2020statistical}, subsampling based stochastic minibatches \citep{gower2019sgd}, and random partition based minibatches \citep{qi2023statistical,gao2023asymptotic}. 
This constitutes the third part of the literature to be reviewed in this work. 

The rest of this article is organized as follows. Section 2 reviews the literature about distributed computing. Section 3 studies various subsampling methods. Section 4 discusses minibatch gradient related techniques. 
The article is then concluded with a brief discussion in Section 5.

\csection{DISTRIBUTED COMPUTING}

\csubsection{Theoretical Framework of Distributed Computing}

Consider a standard statistical learning problem. Assume a total of $N$ observations, where $N$ is notably large. For each observation $i$, we collect a response variable $Y_i\in\mathbb{R}^1$ and its corresponding feature vector $X_i\in\mathbb{R}^p$. The primary goal here is to accurately estimate an unknown parameter $\theta_0\in\mathbb{R}^p$ through a suitably defined loss function. To be specific, define the empirical loss function as $\mL( \theta) = \sum_{i=1}^N \ell (X_i,Y_i;\theta)$. Here, $\ell (X_i,Y_i;\theta)$ represents the loss function for the $i$-th observation.

In conventional scenarios with relatively small $N$, this learning problem could be easily solved using various standard optimization algorithms (e.g., Newton-Raphson method and gradient descent method). Nevertheless, for   
datasets of massive size, implementation of these standard algorithms becomes practically challenging or even infeasible.  
Consider, for example, the classical Newton-Raphson algorithm. 
Let $\wh \theta^{(t)}$ be the estimator derived in the $t$-th iteration. Then, the ($t+1$)-th step estimator is updated as follows:
\beq
\label{eq:newton-raphson}
\wh \theta^{(t+1)} = \wh \theta^{(t)} - \big\{ \ddot{\mL} \big( \wh \theta^{(t)} \big) \big\}^{-1} \dot{\mL}\big( \wh \theta^{(t)} \big),
\eeq
where $\dot{\mL}(\theta)$ and $\ddot{\mL}(\theta)$ represent the 1st- and 2nd-order derivatives of the loss function $\mL(\cdot)$ with respect to $\theta.$ 
With a fixed feature dimension $p$, the computational complexity of the Newton-Raphson algorithm is at least of the order $O(N)$ in each iteration. In the case of large datasets with an exceedingly large $N$, such computation costs could be practically challenging or even infeasible.  To address this issue, various distributed computing methods have been developed. The key idea of distributed computing is to divide a massive dataset into smaller pieces, which can be processed simultaneously across many 
multiple computer machines.

Assume the $N$ samples are distributed across a total of $M$ different local machines and each machine is assigned $n_m$ observations for $1 \leq m \leq M$. 
It follows that $\sum_{m=1}^M n_m = N.$
We then denote the whole  
sample as $\mS_F=\{1,2,...,N\}$ and the sample assigned to the $m$-th local computer as $\mS_{(m)}\subset\mS_F$. Thus, we have $\cup_{m=1}^M \mS_{(m)}=\mS_F$, and $\mS_{(m_1)}\cap \mS_{(m_2)}=\emptyset$  for any ${m_1}\not ={m_2}$, and $|\mS_m| = n_m$. 
Recall that the global loss function is defined as $\mL(\theta) = N^{-1} \sum_{i=1}^N \ell (X_i,Y_i;\theta)$. The averaged local sample size is denoted as $n = M^{-1}\sum_{m=1}^M n_m$.  Define $\wh \theta= \argmin_{\theta} \mL(\theta)$ and $\theta_0 = \argmin_\theta E\big\{ \ell (X_i,Y_i;\theta)\big\}$ as the global estimator and true parameter, respectively. Subsequently, define the local loss function on the $m$-th local computer as $\mL_{(m)} (\theta) = {\color{black} n_m^{-1}} \sum_{i \in \mS_m} \ell(X_i,Y_i;\theta)$. Let $\thetam = \argmin_\theta \mL_{(m)} (\theta)$ be the estimator locally obtained on the $m$-th local computer. Moreover,  denote  $\dot{\ell}(X_i,Y_i;\theta) = \partial \ell(X_i,Y_i;\theta) /\partial \theta \in \mathbb{R}^p$ and $\ddot{\ell}(X_i,Y_i;\theta) = \partial \ell(X_i,Y_i;\theta) / \partial \theta \theta^\top \in \mathbb{R}^{p \times p}$ as  the 1st- and 2nd-order derivatives of $\ell(X_i,Y_i;\theta)$ with respect to $\theta$, respectively. 

\csubsection{One-Shot Methods}
\noindent

For distributed statistical learning, various one-shot (OS) methods have been developed \citep{mcdonald2009efficient,zinkevich2010parallelized,zhang2013communication,rosenblatt2016on,lee2017communication,hector2020doubly}. The basic idea is to calculate some important statistics on each local machine based on the data stored in each local machine in a fully parallel way. Subsequently, they are sent to the central machine, where these statistics are then assembled into one final estimator.
Specifically, each local machine $1 \leq m \leq M$
uses local sample $\mS_{(m)}$ to compute the local estimator $\thetam = \argmin_\theta \mL_{(m)}(\theta)$. Subsequently, the central server collects these local estimates and aggregates them to obtain the final estimator $M^{-1} \sum_{m=1}^M \thetam$, which is denoted as the OS estimator $\thetaoneshot$.

Extensive research has been proposed in this field. For example, \cite{chen2014split} studied the properties of one-shot estimator based on penalized generalized linear regression with smoothly clipped absolute deviation (SCAD) penalty.  \cite{battey2018distributed} proposed high-dimensional one-shot estimators based on Wald tests and Rao's score tests. They extended the classical one-shot estimator from low-dimensional generalized linear regression to high-dimensional sparse scenarios. \cite{lian2018divide} developed a debiased form of one-shot estimator for support vector machines for ultra-high-dimensional data. \cite{TANG2020104567} used the confidence distribution approach to combine bias-corrected lasso-type estimates computed in each local machine in the generalized linear model setting. The one-shot strategy for correlated outcomes is also discussed in the previous literature. One notable work is the distributed and integrated method of moments (DIMM) proposed by \cite{Hector2020a}, which addresses the estimation problem in
a regression setting with high-dimensional correlated outcomes. The key idea is to split all outcomes into blocks of low-dimensional response subvectors, then analyze these blocks in a distributed scheme, and finally combine the block-specific results using a closed-form meta-estimator. By this way, the computational challenges associated with high-dimensional correlated outcomes are alleviated. A generalization of DIMM is further developed in \cite{hector2020doubly}, which doubly divides the data at both the outcome and subject levels to speed up computation. Recently, a distributed empirical likelihood (DEL) method has been proposed to solve the estimation problem for imbalanced local datasets in the framework of integrative analysis  \citep{Zhou2023DistributedEL}.

As one can see, the OS method is easy to implement. It is also communicationally efficient because it requires only one round of communication between the local computers and the central computer (i.e., transferring the local estimates $\thetam$s).  However, many researchers \citep{zhang2013communication,WANG2021107265,wu2023subsampling} have pointed out that a number of critical conditions are necessarily needed by various OS methods to achieve the same asymptotic efficiency as the global estimator $\wh \theta$. The first condition is \emph{uniformity}, implying that the massive data should be distributed across local computers in a relatively uniform manner so that the local sample sizes across different local machines should be approximately equal. The second condition is \emph{randomness}, indicating that the massive data should be distributed across local computers as randomly as possible.
The third condition is \emph{sufficiency}, signifying that the sample size on each local machine should not be too small. To be more precise, it typically requires $n^2 / N \to \infty$ unless some important biased reduction techniques (e.g., jackknifing) have been used \citep{wu2023subsampling}.

However, in real practice, these conditions are often violated to some extent. For example, practitioners rarely distribute large datasets in a completely uniform and random manner. Consequently, understanding how the violation of these conditions affects the statistical performance of the OS estimators becomes a topic of significant interest.
Intuitively, when the uniformity condition is violated, at least one local computer ends up with a relatively tiny sample size. Consequently, the local estimates generated by these local computers may exhibit significantly larger variability or bias than others. When the randomness condition is violated, the local estimates from different local computers could be seriously biased. When the sufficiency condition is violated, the bias of each local estimator $\thetam$ with an order $O(1/n)$ becomes non-negligible as compared with $O(1/\sqrt{N})$, leading to a noticeable bias in the resulting OS estimator $\thetaoneshot$.
In each scenario, $\thetaoneshot$ becomes statistically inefficient or even inconsistent, as rigorously demonstrated by \cite{WANG2021107265}.

To address the challenges posed by the lack of distribution uniformity and randomness, a one-step upgraded pilot (OSUP) estimator was proposed by \cite{WANG2021107265}. The OSUP method comprises several steps and is well-suited for a broad class of models with a likelihood specification. Specifically, to compute the OSUP estimator, a number of $n_0$ pilot samples should be randomly selected from different local computers and then transferred to the central computer. Then the central computer computes a pilot estimator $\widehat{\theta}_{p}$ by maximizing the log-likelihood function $\ell(X_i, Y_i;\theta)$ of all the pilot samples, i.e., $\widehat\theta_{p}=\argmax_\theta \sum_{i\in\mP}\ell(X_i, Y_i;\theta)$, where $\mP$ is the set of pilot samples. The pilot estimator $\widehat{\theta}_{p}$ is $\sqrt{n_0}$-consistent for the target parameter $\theta$. However, due to its smaller sample size (i.e., $n_0 \ll N$), $\widehat{\theta}_{p}$ is not statistically as efficient as the global estimator. To further improve the statistical efficiency, the central computer broadcasts $\widehat{\theta}_{p}$ back to all local computers. Then each local computer considers $\widehat{\theta}_{p}$ as an initial point and computes the 1st and 2nd order derivatives for its local log-likelihood function as
        $\dot\ell_m(\widehat\theta_{p})=\sum_{i\in\mS_{(m)}}\dot\ell(X_i, Y_i;\widehat\theta_{p})$ and
        $\ddot\ell_m(\widehat\theta_{p})=\sum_{i\in\mS_{(m)}}\ddot\ell(X_i, Y_i;\widehat\theta_{p})$.
These derivatives are then communicated to the central computer for summation, i.e.,
$\dot\ell(\widehat\theta_{p})=\sum_{m}\dot\ell_m(\widehat\theta_{p})$ and $\ddot\ell(\widehat\theta_{p})=\sum_{m}\ddot\ell_m(\widehat\theta_{p})$. Based on the summarized derivative information along with the pilot estimate, a novel one-step upgrading is performed by the central computer. This leads to the final OSUP estimator as follows,
\begin{equation}
    \label{eq:osup}
        \widehat\theta_{\text{OSUP}}=\widehat\theta_{p}-\Big\{\frac{1}{N}\ddot\ell(\widehat\theta_{p})\Big\}^{-1}
        \Big\{\frac{1}{N}\dot\ell(\widehat\theta_{p})\Big\}. \nonumber
    \end{equation}
\noindent
Compared with a standard OS estimator, the OSUP estimator incurs an extra computational cost for obtaining the pilot sample. However, the benefits derived from the OSUP method are significant, leading to an estimator with the same statistical efficiency as the global estimator under very mild conditions \citep{WANG2021107265}. A similar one-step estimator is also studied by \cite{huang2019a}. However, the key difference is that the initial estimator used in \cite{huang2019a} is the simple average of all local estimators.

To address the challenges posed by the lack of local sufficient sample size, \cite{wu2023subsampling} developed a jackknife debiased (JDS) estimator to reduce the estimation bias based on the moment estimator. It should be noted that this method was originally proposed for subsampling. However, the key idea is also readily applicable to distributed computing.
To be specific, they first define a jackknife estimator $\color{black} \widehat\theta^{(m)}_{-j}$ for the $m$-th machine as
\[
\widehat\theta^{(m)}_{-j} = \argmin_{\theta} \frac{1}{n-1} \sum_{i \in \mS_{(m)}}^{i \neq j} \ell(X_i,Y_i;\theta).
\]
It could be verified that $\text{Bias}\big(\widehat\theta^{(m)}_{-j}\big)$ approximately equals $\tau/(n-1)$ for some constant $\tau$ {\color{black} \citep{shao1995jackknife}}. Then, $n^{-1} \sum_{j \in \mathcal{S}_m} \text{Bias}\big(\widehat{\theta}^{(m)}_{-j}\big) \approx \tau/(n-1)$ and $E\big(n^{-1} \sum_{j \in \mathcal{S}_m} \widehat{\theta}^{(m)}_{-j} - $  $ \widehat{\theta}^{(m)}\big) \approx \tau/\{n(n-1)\}$. This inspires an estimator for the bias, which is given by $\widehat{\text{Bias}}^{(m)} = (n-1) n^{-1} \sum_{j \in \mathcal{S}_m} \widehat{\theta}^{(m)}_{-j} - (n-1) \widehat{\theta}^{(m)} $. Accordingly,  a bias-corrected estimator for the $m$-th machine could be proposed  as $\thetaJDS^{(m)} = \widehat{\theta}^{(m)} - \widehat{\text{Bias}}^{(m)}.$
Thereafter, $\thetaJDS^{(m)}$s can be further averaged across different $m$. As a consequence, the final JDS estimator could be obtained as
$\thetaJDS = M^{-1} \sum^M_{m = 1} \thetaJDS^{(m)}$. Subsequently, \cite{wu2023subsampling} rigorously verified that $\text{Bias}(\thetaJDS) = O(1/n^2) + O(1/N)$ and the asymptotic variance of $\thetaJDS$ remains the same as that of $\wh \theta$. As a consequence, excellent statistical efficiency can be achieved by $\thetaJDS$ with a very small size $n$. This bias correction method has been theoretically studied for moment estimator; however, the theoretical properties for the estimator computed from a general loss function remain unknown.

\csubsection{Efficient Iterative Approach}
\noindent

To improve the statistical efficiency of one-shot estimators, various distributed iterative methods can be considered. Since distributed computing requires passing messages among multiple computers, naively applying traditional iterative methods to a distributed system often incurs expensive communication costs. Therefore, how to achieve excellent statistical efficiency with well-controlled communication costs becomes the key issue \citep{jordan2018communication}.

To illustrate this point, consider for example extending the iterative Newton-Raphson algorithm \eqref{eq:newton-raphson} to the distributed scenario. Recall that $\mS_{(m)}$ collects the indices of samples allocated to the $m$-th local computer. Given the current estimator $\wh \theta^{(t)}$, we can compute the 1st and 2nd order derivatives of the loss function as
\begin{equation*}
\dot{\mL}\big(\wh \theta^{(t)} \big) = M^{-1} \sum_{m=1}^M \dot{\mL}_{(m)}\big(\wh \theta^{(t)} \big) \quad \textup{and} \quad
\ddot{\mL}\big(\wh \theta^{(t)} \big) = M^{-1} \sum_{m=1}^M \ddot{\mL}_{(m)}\big(\wh \theta^{(t)} \big), 
\end{equation*}
where $\dot{\mL}_{(m)}(\wh \theta^{(t)} ) = \sum_{i \in \mS_{(m)}} \dot{\ell}( X_i,Y_i; \wh \theta^{(t)} ) $ and $\ddot{\mL}_{(m)}(\wh \theta^{(t)} ) = \sum_{i \in \mS_{(m)}} \ddot{\ell} ( X_i,Y_i; \wh \theta^{(t)} )$. Note that $\dot{\mL}_{(m)}(\wh \theta^{(t)} )$ and $\ddot{\mL}_{(m)}(\wh \theta^{(t)} )$ are computed on the $m$-th local computer. They are then transferred to the central computer to update $\wh \theta^{(t+1)}$ according to \eqref{eq:newton-raphson}. As one can see, this is a solution easy to implement but suffers several serious limitations. First, inverting the $(p\times p)$-dimensional Hessian matrix $\ddot{\mL}(\wh \theta^{(t)})$ in the central computer incurs a computation cost with the order $O(p^3)$ for each iteration. Second, transferring the local Hessian matrix $\ddot{\mL}_{(m)}(\wh \theta^{(t)})$ from each local computer to the central computer incurs a communication cost of order $O(p^2)$ for each local computer in each iteration. Thus, this approach leads to high computation and communication costs for high-dimensional data.

To address this issue, various communication-efficient Newton-type methods have been proposed to alleviate high communication costs. One of the underlying key ideas is to avoid Hessian matrix transmission
\citep{shamir2014communication,zhang2015disco,wang2017efficient,wang2018giant,crane2019dingo,jordan2018communication,luo2020renewable}. For example, the entire sample Hessian matrix can be approximated using some appropriate local estimators, which are computed on one single computer (e.g., the central computer). Consequently, the communication cost due to transferring the whole sample Hessian matrix between computers can be avoided. One notable work in this regard is \cite{jordan2018communication}. Motivated from the Taylor series expansion of $\mL(\theta)$, \cite{jordan2018communication} defined
a surrogate loss function as $\widetilde{\mL}(\theta) = \mL_{(1)}(\theta) - \theta^{\top}\left\{ \dot\mL_{(1)}(\overline{\theta})-\dot\mL(\overline{\theta})\right\}$,
where $\overline{\theta}$ denotes any initial estimator of $\theta$. Then, based on the surrogate loss function, the updating formula of Newton's method is modified as $\widehat{\theta}^{(t+1)}= \widehat{\theta}^{(t)}- \{\ddot\mL_{(1)}(\widehat{\theta}^{(t)})\}^{-1} \dot\mL(\widehat{\theta}^{(t)})$,
where $\ddot\mL_{(1)}(\widehat{\theta}^{(t)})$ is the local Hessian matrix computed on the 1st local computer. Thus no Hessian matrix communication is needed for each iteration.

The implementation of communication-efficient Newton-type methods significantly reduces communication costs. However, when dealing with high-dimensional data, computing the inverse of the Hessian matrix remains to be a computationally expensive problem. To further improve computation efficiency, various methods avoiding matrix inverse calculation have been proposed.
The first type is distributed (stochastic) gradient descent algorithms \citep{goyal2017accurate,lin2018distributed,qu2019accelerated,su2019securing,li2022statistical,chen2022first}, which compute only the 1st order derivatives of the loss function (i.e., gradients).
To be specific, in distributed gradient descent methods, each local computer first receives the current parameter estimator from the central computer. Once the current parameter estimator is received, each local computer calculates its own gradient and sends it back to the central computer. Lastly, the central computer aggregates the local gradients and updates the parameter estimator as
\beq
\label{eq:dgd}
\wh \theta^{(t+1)} = \wh \theta^{(t)} - \alpha_t M^{-1} \sum_{m=1}^M \dot{\mL}_{(m)}(\wh \theta^{(t)}),
\eeq
where $\alpha_t>0$ represents the learning rate.
To further reduce the communication cost in \eqref{eq:dgd}, a local (stochastic) gradient descent algorithm is proposed \citep{stich2019local,woodworth2020local}. The key idea is to run the (stochastic) gradient descent algorithm independently and locally on different local computers in a fully parallel way. Subsequently, the local estimators are transferred to the central computer and then iteratively updated to form the final estimator.

However, for those distributed gradient descent algorithms, a large number of iterations are typically required for numerical convergence,  and the choice of hyperparameters (e.g., $\alpha_t$) is critically important and also difficult \citep{yingqiu2021}. To address this problem, various quasi-Newton methods in a distributed manner have been developed  \citep{chen2014bfgs,Eisen2017,Lee2018,Soori2020dqn,wu2023quasi}. The key idea of distributed quasi-Newton methods is to approximate the Hessian inverse in each iteration without actually inverting the matrix \citep{davidon1991variable,goldfarb1970family}.
The communication cost of these methods could have orders as low as $O(p)$ in each iteration. In the meanwhile, the convergence rate of distributed quasi-Newton methods is superlinear, surpassing the linear convergence of distributed gradient descent methods \citep{broyden1973local}.

As an important method along this direction,
\cite{wu2023quasi} developed a $K$-stage distributed quasi-Newton method. Specifically, \cite{wu2023quasi} started with the following one-stage distributed quasi-Newton method
\beq
\label{eq:one-step}
\thetaos = \thetap - M^{-1} \sum_{m=1}^M  \Big\{ H_{(m,0)}  \dot{\mL}(\thetap) \Big\}.
\eeq
Here $\thetap$ is a suitable initial estimator, such as the OS estimator $\thetaoneshot$. $H_{(m,0)}$ represents the local inverse Hessian estimator obtained by each local computer after implementing the quasi-Newton algorithm.
Next, the local computer sends $H_{(m,0)}  \dot{\mL}(\thetap)$ as a whole to the central computer. Therefore, the communication cost is of the order $O(p)$. \cite{wu2023quasi} has verified that the optimal statistical efficiency can be achieved by the one-stage distributed quasi-Newton estimator as long as $N(\log p)^4/n^4 \to 0$. This condition can be further relaxed by allowing for multi-stage updating. This leads to the $K$-stage distributed quasi-Newton estimator, which is extremely efficient both computationally and communicationally.

\csubsection{Distributed Quantile Regression}

Quantile regression is an important class of regression methods for its robustness against heavy-tailed distributed responses and outliers \citep{koenker1978regression,koenker2005}. Its applications span various disciplines, including agriculture \citep{kostov2013quantile}, climate change \citep{reich2011bayesian}, health studies \citep{alhamzawi2018bayesian}, house pricing \citep{chen2013bayesian}, and others \citep{xu2017estimation,zhong2022estimation}. With the availability of large-scale datasets, extensive research has been dedicated to distributed estimation and inference for quantile regression.
For instance, \cite{yang2013quantile} proposed a subspace preserving sampling technique for quantile regression on massive datasets. However, their method suffers from the problem of statistical inefficiency. Later, \cite{xu2020block} developed a block average approach, employing the one-shot strategy by averaging estimators derived from each local computer. 

To guarantee statistical efficiency, researchers address the challenges of distributed quantile regression by proposing various loss functions and iterative algorithms. For instance, \cite{volgushev2019distributed} proposed a two-step quantile projection algorithm, incorporating valid statistical inference. In the initial step, conditional quantile functions are estimated at different levels. Subsequently, a quantile regression process is constructed through projection. \cite{chen2019quantile} presented a computationally efficient method, which involves multiple rounds of aggregations. 
After limited $q$ iterations, the authors show that the statistical efficiency of the final estimator becomes the same as the one computed on the whole data. 
In a related work, \cite{chen2020distributed} studied the linear regression problem with heavy-tailed noises. Since the quantile regression loss function is a non-smooth function, the authors established a connection between quantile regression and ordinary linear regression by transforming the response. This results in a distributed estimator that is efficient in both computation and communication. 
Instead of dealing with conventional smoothing functions, \cite{hu2021distributed} extended the communication-efficient surrogate likelihood method proposed by \cite{jordan2018communication}. The authors constructed a surrogate loss function and established the consistency and asymptotic normality of the proposed methodology. In a recent development, \cite{tan2022communication} developed a double-smoothing approach to the local and global objective functions of quantile regression.

In a recent work of \cite{pan2022note}, the authors proposed a one-step approach that is efficient both communicationally and statistically. Notably, the derived estimator is robust against data distribution heterogeneity across local computers. Specifically, assume there are $N$ observations indexed by $i = 1,\dots, N$. The response $Y_i$ and the $p$-dimensional predictor $X_i$ follow the standard $\tau$-th quantile regression model $Y_i = X_i^\top\beta_{\tau}+\varepsilon_i$,
where $\beta_{\tau}$ is the associated regression coefficient vector and $\tau\in(0,1)$. Additionally, $\varepsilon_i$ is the error term satisfying $P(\varepsilon_i\leq 0|X_i)=\tau$. The standard check loss function can be constructed as $\mathcal{L}(\bbeta)=\sum_{i=1}^{N}\rho_{\tau}(Y_i-X_i^\top\bbeta)$,
where $\rho_{\tau}(\mu)=\mu\{\tau-I(\mu\leq0)\}$ represents the check function and $I(\cdot)$ the indicator function. Consequently, the standard estimator of $\beta_{\tau}$ can be obtained by $\hat\beta_{\tau}=\mathop{\arg\min}_{\beta}\mathcal{L}(\beta)$.
Assume that a pilot sample (i.e., indexed by $\mathcal{Q}$) is derived across $K$ local computers. The sample size of the pilot sample is $n$ satisfying $n/N\rightarrow 0$. As a result, a pilot estimator can be obtained by $\hat\beta^{\mathcal{Q}}_{\tau}=\mathop{\arg\min}_{\beta}\sum_{i\in\mathcal{Q}}\rho_{\tau}(Y_i-X_i^\top\beta)$, which is $\sqrt{n}$-consistent. The one-step estimator proposed by \cite{pan2022note} is derived as
\begin{equation}\label{estimator}
\hat\beta^{(1)}_{\tau}=\hat\beta^{\mathcal{Q}}_{\tau}+\frac{1}{\hat{f}(0)}\left(\sum_{i=1}^{N}X_iX_i^{\top}\right)^{-1}
\left[\sum_{i=1}^{N}X_i\Big\{\tau-I(\hat\varepsilon_i\leq0)\Big\}\right],
\end{equation}
where $\hat f(\cdot)$ is a kernel density estimator and $\hat\varepsilon_i$ is the residual. It can be proved that $\hat\beta^{(1)}_{\tau}$ is $\sqrt{N}$-consistent and asymptotically normal, regardless of how the raw data are distributed across local computers.

\csubsection{Distributed Logistic Regression with Rare Events Data}

{\color{black}
The rare events problem in this subsection refers to a binary data classification problem, where one class (often assumed to be the negative class) has a much greater number of instances than the other class (often assumed to be the positive class). 
Rare events data are prevalent in scientific fields and applications.
The rare events data examples include but are not limited to drug discovery \citep{zhu2006lago, korkmaz2020deep}, software defects \citep{richardson2013infection}, and rare disease diagnosis \citep{zhao2018framework,zhuang2019care}.}
In traditional statistical theory, one often assumes that the probability of any type of event to happen is fixed.
However, for rare events, it is more appropriate to assume that the positive class probability should decay towards zero at an appropriate rate as the total sample size increases \citep{wang2020logistic}.
In this regard, \cite{wang2020logistic} constructed a novel theoretical framework to accurately describe rare events data.
Under this theoretical framework, it was demonstrated that the convergence rate of the global maximum likelihood estimator (MLE) is mainly determined by the sample size of the positive class instead of the total sample size.
This implies a considerably slower convergence rate than that of the classical cases. 
It seems that limited attempts have been made for the rigorous asymptotic theory for distributed classification problems with rare event data.
This motivates \cite{li2023distributed} to develop a novel distributed logistic regression method with solid statistical theory support for massive rare event data.

More specifically, assume there are a total of $N$ observations indexed by $1 \le i \le N$.
The $i$-th observation is denoted as $\big(X_i, Y_i\big)$, where $X_i \in \mathbb{R}^{p}$ is a $p$-dimensional covariate and $Y_i \in\{0,1\}$ is the binary response.
{\color{black} Let $N_1=\sum_{i=1}^{N} Y_i$ be the total number of positive instances.}
To model their regression relationship, the following logistic regression model is considered
\beqr
\label{eq: logistic}
P\big(Y_i=1 \mid X_i\big) =p_i(\alpha, \beta)= \frac{\exp(\alpha+X_i^\top \beta)}{1+\exp(\alpha+X_i^\top \beta)}, \nonumber
\eeqr
where $\alpha \in \mathbb{R}$ is the intercept and $\beta \in \mathbb{R}^p$ is the slope parameter.
To reflect the asymptotic behavior, two important assumptions should be imposed \citep{wang2020logistic}.
First, the percentage of positive instances should be extremely small.
Statistically, the positive response probability is specified to converge towards 0 as the total sample size $N \to \infty$.
Rewrite $\alpha$ as $\alpha_{N}$. 
This leads to $\alpha_N \to -\infty$ as $N \to \infty$.
Second, the total number of positive instances should diverge to infinity. 
Otherwise, the parameters of interest cannot be estimated consistently.
Mathematically, it follows that $E(N_1) \approx N \exp(\alpha_N) E\{\exp(X_i^\top \beta)\}$ as $N\to \infty$. 
This suggests that $\alpha_N\to-\infty$ and $\alpha_N+\log N\to\infty$ as $N\to \infty$ \citep{wang2020logistic,li2023distributed}.

Assume a distributed computation system with one central computer and a total of $K$ local computers indexed by $1\le k\le K$.
\cite{li2023distributed} first discussed two different data distribution strategies. 
They are RANDOM and COPY strategies, respectively.
Specifically, the RANDOM strategy is used to randomly distribute the full data to each local computer with approximately equal sizes.
For the COPY strategy, all the positive instances are copied to every local computer.
In contrast, the negative instances are randomly distributed on different local computers.
Next, \cite{li2023distributed} studied three types of objective functions. 
They are, respectively,
\beqr
\mL_{{\rm R},k} \big(\theta\big) &=& \sum_{i=1}^{N} a_i^{(k)} \Big\{Y_i \log p_i\big(\alpha_N,\beta\big) + \big(1-Y_i\big)\log \big(1-p_i(\alpha_N,\beta) \big)\Big\}, \nonumber \\
\mL_{{\rm US},k}(\theta) &=&
\sum_{i=1}^N \Big\{Y_i \log p_i\big(\alpha_N,\beta\big) + \big(1-Y_i\big)a_i^{(k)}\log \big(1-p_i(\alpha_N,\beta) \big)\Big\}, \nonumber \\
\mL_{{\rm IPW},k}(\theta) &=&
\sum_{i=1}^N \Big\{Y_i \log p_i\big(\alpha_N,\beta\big) + K \big(1-Y_i\big) a_i^{(k)}\log \big(1-p_i(\alpha_N,\beta) \big) \Big\}. \nonumber
\eeqr
Here $P(a_i^{(k)}=1)=1/K$ and $a_i^{(k)}=1$ if the $i$-th observation is randomly distributed to $k$-th local computer.
Local estimators can be computed based on the local data according to different loss functions.
The local estimators are then averaged by the central computer in the last step \citep{zhang2013communication,chang2017divide}.
This leads to three distributed estimators (i.e., $\wh{\theta}_{\rm RMLE}$, $\wh\theta_{\rm US}$ and $\wh\theta_{\rm IPW}$).
Under some regularity conditions, 
\cite{li2023distributed} found that the COPY strategy together with the inverse probability weighted objective function $\mL_{{\rm IPW},k}(\theta)$ seems to be the best choice.

\csubsection{Decentralized Distributed Computing}

The distributed computing methods outlined earlier share a common characteristic. That is the requirement of a central computer, which is responsible for communicating with every local computer. 
Such type of architecture is easy to implement. However, it suffers from several serious limitations.
First, this centralized network structure is extremely fragile. If the central computer stops working, the entire network stops. Second, there exists the issue of privacy disclosure for the centralized structure. This is because if the central machine is attacked, the attacker is given the chance to communicate with every local computer. Third, a centralized network structure has a high requirement for network bandwidth, since the central machine should communicate with numerous local computers \citep{bellet2018personalized,li2020blockchain}.

To fix these problems, a number of researchers advocate the idea of fully decentralized distributed computing, which is also called decentralized federated learning (DFL) \citep{yuan2016convergence}. The key feature of DFL is that there is no central computer involved for model training and communication. Different local computers are directly connected through a sophisticated communication network. All computation-related communications should occur only between network-connected individual local computers. 
To be specific, define $\wh \theta^{(t,m)}$ to be the $t$-th estimator obtained on the $m$-th local computer, and the updating formula of the decentralized federated learning algorithm is given by
\beq
\label{eq:ngd}
\widehat{\theta}^{(t+1,m)}=\widetilde{\theta}^{(t,m)} -\alpha \dot{{\mathcal{L}}}_{(m)} \left(\widetilde{\theta}^{(t,m)}\right).
\eeq
Here, $\wt \theta^{(t,m)}$  is the neighborhood-averaged estimator obtained in the $t$-th iteration for the $m$-th local computer. 
Numerous studies have investigated the numerical convergence properties of the method \citep{blot2016gossip,nedic2017achieving,lian2017can,vanhaesebrouck2017decentralized,tang2018d,lalitha2018fully}. It has been demonstrated that the algorithm can achieve a linear convergence rate even with data heterogeneity \citep{richards2020decentralised,savazzi2020federated}. To achieve this nice theoretical property, several stringent conditions have been assumed about the network structure in the past literature. The most typical assumption is that the transition matrix determined by the network structure should be doubly stochastic \citep{yuan2016convergence,tang2018d}. Unfortunately, this is an assumption that can hardly be satisfied in real practice.

To relax this stringent condition, \cite{wu2023network} developed a novel methodology for DFL, which 
only requires the network structure to be weakly balanced. 
To be specific, take a linear regression as an example. Assume there are a total of $M$ local computers, which are connected by a communication network.   
The adjacency matrix of the network is defined as $A = (a_{ij}) \in \mbR^{M \times M}$, and the corresponding weighting matrix is defined as $W=(w_{ij}) \in \mbR^{M \times M}$ with $w_{ij} = a_{ij} / d_{m}$, where $d_m = \sum_j a_{ij}$ represents the in-degree of $A$. 
Then algorithm \ref{eq:ngd} could be rewritten as
\begin{equation}\label{eq:NGD}
\widehat{\theta}^{(t+1,m)}=\left(I_p-\alpha \widehat{\Sigma}_{x x}^{(m)}\right)\Big(\sum_{k=1}^{M} w_{m k} \widehat{\theta}^{(t,k)}\Big)+\alpha \widehat{\Sigma}_{x y}^{(m)},
\end{equation}
where $I_p \in \mbR^{p \times p}$ is an identity matrix, $\wh \Sigma_{xx}^{(m)} = \sum_{i \in \mS_{(m)}} X_i X_i^\top /n  \in \mbR^{p \times p}$, and $\wh \Sigma_{xy}^{(m)} = \sum_{i \in \mS_{(m)}} X_i Y_i / n  \in \mbR^{p}$.
Next, define 
$\widehat{\theta}^{*(t)}= \big(\hat{\theta}^{(t,1) \top}, \dots, \hat{\theta}^{(t,M) \top}\big)^{\top} \in \mathbb{R}^{Mp}, \  \widehat{\Sigma}_{x y}^{*}=\big(\widehat{\Sigma}_{x y}^{(1) \top},$\quad$ \dots, \widehat{\Sigma}_{x y}^{(M) \top}\big)^{\top} \in \mathbb{R}^{Mp},$ and  $\Delta^* = \diag \big\{I_p - \alpha \widehat{\Sigma}^{(1)}_{xx}, \dots, I_p - \alpha \widehat{\Sigma}^{(M)}_{xx} \big\}$. This leads to a matrix form of (\ref{eq:NGD}) as 
\[
\widehat{\theta}^{*(t+1)}  =\Delta^{*}(W \otimes  I_p) \widehat{\theta}^{*(t)} + \alpha \wh \Sigma^*_{xy},
\]
where $\otimes$ stands for the Kronecker product. 
Assume the stable solution of this system 
(denoted by $\wh{\theta}^{*}$) exists, it follows then
$
\widehat{\theta}^{*}  = \alpha \Big\{ I_p - \Delta^{*}(W \otimes  I_p)  \Big\}^{-1}  \wh \Sigma^*_{xy}. 
$
Theoretically, \cite{wu2023network} proved that the statistical efficiency of the DFL is determined by three factors: (1) the learning rate, (2) the network structure, and (3) the data distribution pattern.  
The optimal statistical efficiency can be guaranteed if the learning rate is relatively small and the network structure is relatively balanced, even if data are distributed heterogeneously.

There has been extensive research to further extend the classical DFL algorithm from different perspectives. To address the issue of insufficient labels, \cite{gao2019hhhfl} developed a heterogeneous horizontal federated learning framework. To enhance network security, \cite{chen2022decentralized} introduced a decentralized federated learning algorithm that meets differential privacy requirements.  The key idea is that each client adds random noise to their parameter estimators before communication.  The classical DFL algorithm suffers from high communication costs and low convergence rates in non-convex situations. To fix this problem, \cite{nadiradze2021asynchronous} proposed an asynchronous decentralized federated learning method. A novel DFL framework was developed by \cite{liu2022general}, which optimally balances communication efficiency and statistical efficiency. 
\cite{liu2022fast} further proposed a decentralized surrogate median regression method for non-smooth sparse problems. 
For valid statistical inference in DFL, \cite{gu2023decentralized}
studied communication-efficient $M$-estimation. To achieve optimal efficiency, they proposed a one-step DFL estimation method that allows a relatively large number of clients.

\csubsection{Distributed Statistical Inference} 

In addition to estimation, other statistical inference tools, such as hypothesis testing and confidence intervals, also play a crucial role in scientific research and data analysis. 
These inference tools allow researchers to quantify the uncertainty of the estimators obtained from the data, and then help practitioners interpret the results appropriately \citep{casella2002statistical}. In the above discussion of this section, various distributed estimation methods have been introduced. 
Most of them have proven that the distributed estimator can be statistically as efficient as the global estimator under certain conditions \citep{wang2017efficient,jordan2018communication,volgushev2019distributed,wang2019distributed, zhu2021least,fan2021communication, pan2022note}. 
Consequently, the distributed estimators generally share the same asymptotic distribution as the global estimator.
This means that if one can consistently estimate the asymptotic covariance matrix, then asymptotically valid statistical inference can be directly conducted. 
Then the key issue becomes how to estimate the asymptotic covariance matrix consistently and distributedly.
To this end, various plug-in approaches have been widely adopted for various models. 
For example, general $M$-estimation problems with smoothed loss \citep{jordan2018communication}, quantile regression models \citep{pan2022note}, support vector machine \citep{wang2019distributed}, and various debiased estimators for high-dimensional models \citep{fan2019distributed, tu2023distributed}.

However, if the asymptotic covariance of an estimator is too complex, it can be very challenging to construct the corresponding estimator analytically.
In this case, bootstrap provides a more directed inference approach \citep{shao1995jackknife, efron1981jackknife}. Nevertheless, bootstrap usually requires multiple resampling procedures over the whole dataset. This is computationally too expensive to be acceptable, especially for massive datasets. 
To solve this problem, researchers developed some more computationally feasible bootstrap methods. 
Among them, \cite{ArielKleiner2014ASB} introduced a method called the bag of little bootstraps (BLB). 
The BLB method first divides the whole sample into multiple subsets. It then computes multiple repeated estimates of the estimator (or related statistics) based on the inflated resamples.
Finally, an averaging step is applied to aggregate these estimates.
Taking a similar approach, \cite{sengupta2016subsampled} further proposed the method of subsampled double bootstrap (SDB). 
Instead of directly dividing the whole sample into the disjoint subsets, the SDB method selects multiple subsets from the whole dataset by sampling with replacement. 
The computational efficiency and inferential reliability of these subsample-based bootstrap methods depend on various hyperparameters, such as the subsample size and the number of replicates. 
To address this issue, \cite{ma2023optimal} developed an interesting approach for selecting the optimal hyperparameters for the subsample-based bootstrap methods, including the BLB and the SDB.

However, the above bootstrap variants generally require that the involved estimator (or statistic) has a weighted subsample representation. This may not be true for some general statistics. 
For example, the class of symmetric statistics considered in \cite{chenandpeng2021distributed}, which includes the $U$-statistics as an important example.
\cite{chenandpeng2021distributed} investigated the theoretical properties of the one-shot averaging type distributed statistics for both the degenerate and non-degenerate cases. 
For inference purposes, they developed a distributed bootstrap procedure, where no technique such as inflated resampling technique in the BLB method of \cite{ArielKleiner2014ASB} is required.
To further ease the computational burden, \cite{chenandpeng2021distributed} proposed a pseudo-distributed bootstrap (PDB) procedure.
The consistency of the PDB procedure has also been theoretically proved and numerically validated.

\csection{SUBSAMPLING MODELS}

\csubsection{Sequential Addressing Subsampling}

Note that subsampling technique is closely related to the idea of bootstrap \citep{BradleyEfron1979BootstrapMA,bickel1981some}. However, the classical full size bootstrap is often computationally too expensive for massive data analysis. A practical solution in this regard is to repeatedly generate subsamples of small sizes for parameter estimation and statistical inference. Consequently, various subsampling methods are proposed. These methods include, but are not limited to, the $m$ out of $n$ bootstrap \citep{PeterJBickel1997Re}, the bag of little bootstrap \citep{ArielKleiner2014ASB}, the subsampling double bootstrap \textcolor{black}{\citep{sengupta2016subsampled}}, the distributed bootstrap \citep{chenandpeng2021distributed}, the optimal subsampling bootstrap \citep{ma2023optimal}, and possibly others. These methods are particularly useful for the situation, where the data is small enough to be comfortably placed on one single hard drive but large enough so that it cannot be fully loaded into the computer memory as a whole.

When computational resources are limited, an alternative way of subsampling is to comprehensively utilize both the computer memory and the hard drive. 
Some early literature has proposed out-of-core sampling methods, which obtain samples by randomly accessing data points on the hard drive without loading the whole data file in advance \citep{JeffreyScottVitter1985RandomSW, KimHungLi1994ReservoirsamplingAO}. However, due to the hardware limitations at that time, such out-of-core sampling methods have not been tested on massive datasets. 
When dealing with massive datasets, the time cost is of the most critical concern. The time required to sample a single data point from the hard drive often exceeds that of in-memory sampling \citep{SSuwandarathna2007IncreasingHD}. This time cost comprises two main components. They are, respectively, the {\it addressing cost} associated with identifying the target data point on the hard drive and the {\it I/O cost} associated with reading the target data point into memory. 
It is then referred to as the {\it hard drive sampling cost} (HDSC), representing the time needed to fetch a specific data point from the hard disk into computer memory \citep{pan2023sequential}. To reduce the HDSC for massive data, \cite{pan2023sequential} developed a computationally efficient method known as {\it sequential addressing subsampling} (SAS). This method involves a two-step process. They are, respectively, a random shuffling operation aiming at randomly sorting the raw data and a sequential sampling step for obtaining the desired subsamples. It is noteworthy that the random addressing operation, a crucial component of obtaining a subsample, is only performed once. The subsample obtained through this process is referred to as the SAS subsample, and various statistics can be constructed and theoretically studied using these SAS subsamples.

The subsampling method in \cite{pan2023sequential} provides a promising solution to accelerate the subsampling process on the hard drive. Moreover, their SAS method has been proven to be a robust tool for making statistical inference on massive datasets. 
Consider the sample mean as a concrete example to illustrate the theoretical findings based on the SAS method. Assume a set of $N$ samples represented by $X_1,\dots,X_N$ with mean a $\mu$ and variance $\sigma^2$. Additionally, assume that  $E(X_i-\mu)^4=\gamma\sigma^4$. Let $\{X_k,X_{k+1},\dots,X_{k+n-1}\}$ be the $k$-th subsample with a sample size of $n$, where the sample mean is defined as $\overline X_{k}=n^{-1}\sum_{i=k}^{k+n-1} X_i$. In practice, assume that $B$ sequential subsamples are obtained using the SAS method. Denote the corresponding sample means of these $B$ sequential subsample as $\{\overline{X}_{(1)},\dots,\overline{X}_{(b)},\dots,\overline{X}_{(B)}\}$. Define the sample mean of interest as 
$\overline{\overline{X}}_B=B^{-1}\sum_{b=1}^B \overline{X}_{(b)}$.
It is easily to see that $E(\overline{\overline{X}}_B) = \mu$. Given the assumptions that (1)  $n\rightarrow\infty$ and $n/N\rightarrow 0$ as $N\rightarrow\infty$; (2) $B/N\rightarrow 0$ and $nB = O(N)$ as $N \to \infty$, the variance of $\overline{\overline{X}}_B$ can be presented as  
\begin{equation*}
\var(\overline{\overline X}_B)=\sigma^2\bigg(\frac{1}{nB}+\frac{1}{N}\bigg)\big\{1+o(1)\big\}.
\end{equation*}
On one hand, the term $\sigma^2/N$ is associated with the overall sample and cannot be eliminated by subsampling. On the other hand, the term $\sigma^2/(nB)$ can be reduced by increasing the subsample size $n$ or the number of subsamples $B$. To perform automatic inference, the standard error of $\overline{\overline{X}}_B$ is proposed as 
\begin{equation*}
\wh{\mbox{SE}}^2(\overline{\overline X}_B) = \frac{n}{B-1}\Big(\frac{1}{nB}+\frac{1}{N}\Big) \sum_{b=1}^B \Big(\overline{X}_{(b)}-\overline{\overline{X}}_{B}\Big)^2,
\end{equation*}
where the theoretical results can be found in \cite{pan2023sequential}. In summary, the SAS method is time-saving in terms of HDSC as well as useful for automatic statistical inferences.

\csubsection{Subsampling-based Estimation Methods}

In the previous subsection, we have thoroughly reviewed various subsampling methods and discussed them from the perspective of computational efficiency. In the meanwhile, how to estimate the parameters of interest with the best statistical efficiency is also a crucial concern. To formulate this problem, let  $\mathcal{F}=\{i:1\leq i \leq N\}$  represent an index set for an extremely large dataset with sample size $N$. Let $Y_i $ be the response associated with the $i$-th subject and $X_i=(X_{ij})\in \mathbb{R}^p$ be the corresponding $p$-dimensional feature vector. Define $\pi_i$ to be the sampling probability for each sample $1\leq i\leq N$. A random subsample of size $n$ can be drawn (with replacement) with $\pi=(\pi_1,...,\pi_N)^{\top}\in \mathbb{R}^N$. Then the key research question here is how to define $\pi_i$s appropriately, so that a small but representative subsample can be obtained for downstream statistical models.

Various subsampling methods have been proposed to address this concern. For the linear regression model, the algorithm leveraging methods have been extensively discussed, utilizing the empirical statistical leverage scores of the input covariate matrix to define the sampling probabilities \citep{drineas2006sampling,drineas2011faster,mahoney2011randomized}. \cite{ma2014statistical} further provided an effective framework to evaluate the statistical properties of parameter estimation in these algorithmic leveraging methods. \cite{wang2019information} introduced the information-based optimal subdata selection (IBOSS) method, which can deterministically identify a subsample with the maximum information matrix under the $D$-optimality criterion. The IBOSS approach is further extended to a divide-and-conquer setting by \cite{wang2019divide}. For binary logistic regression, \cite{wang2018optimal} proposed the optimal subsampling method motivated by the $A$-optimality criterion (OSMAC) by minimizing the asymptotic mean squared error (MSE) of the subsample estimator to design the subsampling probability.
The OSMAC method can be enhanced by incorporating unweighted objective functions and Poisson subsampling, resulting in improved efficiency \citep{wang2019more}.
 Furthermore, the applicability of OSMAC is extended to various classes of models, including  multi-class logistic regression \citep{yao2019optimal}, generalized linear models \citep{ai2021optimal}, quantile regression \citep{wang2021optimal} and quasi-likelihood \citep{yu2022optimal}.

Although the above methods are demonstrated to be statistically efficient, computing $\pi_i$ for the entire dataset poses a substantial computational challenge, especially when the data size $N$ is extremely large. Take the OSMAC method \citep{wang2018optimal} as an example. For the OSMAC method, determining the optimal subsampling probabilities involves a computational complexity of $O(Np)$. As a result, this optimal subsampling algorithm becomes computationally expensive when dealing with a very large sample size $N$. To address this challenge, the repeated subsampling method can be adopted. The key idea of repeated subsampling is to draw a subsample with uniform probability (i.e., $\pi_i=1/N$) while operating the subsampling step repeatedly. Using the uniform probability in the subsampling process eliminates the need to compute probabilities for the entire dataset in advance. Therefore the computation cost is significantly reduced. Through repeated subsampling, the selected data approximates the whole dataset. Note that the cost associated with subsampling cannot be negligible, particularly true for the repeated subsampling methods. Nevertheless, thanks to the SAS method of \textcolor{black}{\cite{pan2023sequential}}, the hard drive sampling cost can be significantly reduced.

Based on repeated subsampling, a variety of statistical models have been developed. Here we introduce the sequential one-step (SOS) estimator for generalized linear models \citep{Wang2022SequentialOE} for example. Assume the whole dataset has been randomly distributed on the hard drive and the SAS method is used to obtain each subdata. Assume the subsampling is repeated for $K$ times. In the $k$-th subsampling with $1\leq k\leq K$, denote $\mS_k$ to be the indices of selected observations in the whole dataset. Based on $\mS_k$, the SOS estimator is computed as follows. First, we need to calculate an initial estimator $\overline\beta_{1}$ based on $\mS_{1}$. This initial estimator could, for instance, be a maximum likelihood estimator (MLE) of the generalized linear regression models. Assume $\overline\beta_{k}$ to be the current estimator in the $k$-th step. Subsequently, in the $(k+1)$-th subsampling step, a new SAS subsample $\mS_{k+1}$ is obtained. Then a one-step update is performed based on $\overline\beta_{k}$ to obtain the one-step updated estimator, i.e., $\widehat\beta_{k+1}=\overline\beta_{k}-\big\{\ddot\ell_{\mS_{k+1}}(\overline\beta_{k})\big\}^{-1}\dot\ell_{\mS_{k+1}}(\overline\beta_{k})$,
where $\dot\ell_{\mS_{k+1}}(\overline\beta_{k})$ and $\ddot\ell_{\mS_{k+1}}(\overline\beta_{k})$ denote the 1st and 2nd order derivatives of the likelihood function based on the $(k+1)$-th subsample, respectively. Next, the SOS subsampling estimator for the $(k+1)$-th step is computed as
$\overline\beta_{k+1}=\big\{k\overline\beta_{k}+\widehat\beta_{k+1}\big\}/(k+1) =\sum_{l=1}^{k+1}\widehat\beta_{l}/(k+1)$. The corresponding estimator obtained in the last $K$-th step is the final SOS estimator, i.e., $\widehat{\beta}^{\text{SOS}}=\overline\beta_{K}$. It is noteworthy that the SOS method represents an extension of the classical one-step estimator \textcolor{black}{\citep{Shao2003,zou2008one}} but within the context of subsampling. The theoretical properties of the SOS estimator are also established in \cite{Wang2022SequentialOE}, demonstrating that both the bias and variance of the SOS estimator decrease as the number of sampling iterations $K$ increases.

\csubsection{Subsample Feature Screening}

Feature screening plays a critically important role for ultrahigh dimensional data analysis. 
Extensive literature has been developed along this direction. Since \cite{fan2008sure} introduced the seminal work of sure independence screening (SIS), a large amount of follow-up research has been inspired. The key idea of SIS is to rank and then select important features by certain appropriately
defined correlation measures. For example, under a linear regression model setup and assuming some appropriate regularity conditions, \cite{fan2008sure} showed that the top features selected according to marginal sample correlation coefficients are screening consistent. In other words, the selected top features are assured to asymptotically cover the underlying low dimensional true model with probability tending to one. 
\cite{wang2009forward} further improved SIS by the method of forward regression for a significantly improved finite-sample performance. \cite{li2012feature} proposed a distance correlation based independent screening method (DC-SIS) so that variable screening can be conducted in a model free manner. 
Recently, a distributed feature selection method has been developed by \cite{li2020distributed} for massive data analysis.  

\cite{zhu2022feature} proposed a novel subsampling-based feature selection method for large datasets with ultrahigh dimensional features.  
They consider a classical linear regression model as 
$
Y_{i}=X_{i}^\top \beta+\varepsilon_{i}
$
\citep{fan2008sure},
where $\beta \in \mathbb{R}^{p}$ is regression parameters, $\varepsilon_{i}$ is the independent noise term with $\operatorname{var}\left(\varepsilon_{i}\right)=\sigma^{2}$. 
Assume a total of $B$ subsamples with size $n$, which are denoted by 
$
\mS_{(b)} \subset \mS_{F} = \big\{ 1,2,\dots,N \big\}
$
with $|\mS_{(b)}| = n$.
Define $\mathbb{X}_{(b)} = (X_i:i\in\S_{(b)}) \in \mbR^{n \times p}$ as the subsampled design matrix and $\mathbb{Y}_{(b)}=(Y_i: i \in \mS_{(b)} ) \in \mbR^{n}$ as the associate  response vector.
Define a candidate model as $\mathcal{M}=\{j_{1}, \ldots, j_{m}\}$ with $1 \leq j_k \leq p$ for every $1 \leq k \leq K$. Define the design matrix associate with $\mathcal{M}$ as 
$
\mathbb{X}_{(b)}^{(\mathcal{M})} \in \mbR^{n \times |\mM|}
$.
Then, the $R$-Squared statistic to be computed from the $b$-th subsample for the model $\mathcal{M}$ is given by   
\[
R^{2}_{(b)} \left(\mathcal{M} \right)=\Big(\mathbb{X}_{(b)}^{\mathcal{M} \top} \mathbb{Y}_{(b)}\Big)^\top\Big(\mathbb{X}_{(b)}^{\mathcal{M} \top} \mathbb{X}_{(b)}^{\mathcal{M}}\Big)^{-1}\Big(\mathbb{X}_{(b)}^{\mathcal{M} \top} \mathbb{Y}_{(b)}\Big) \Big\|\mathbb{Y}_{(b)}-\overline{\mathbb{Y}}_{(b)}\Big\|^{-2},
\]
where $\overline{\mathbb{Y}}_{(b)}=n^{-1} \mathbf{1}^\top \mathbb{Y}_{(b)}$.  
Thereafter, a one-shot type statistic can be assembled as 
$R^{2}_{\mathrm{OS}}\left( \mathcal{M} \right) = B^{-1} \sum_{b=1}^{B} R^{2}_{(b)}(\mathcal{M}).$ 

The one-shot estimator $R^{2}_{\mathrm{OS}}\left( \mathcal{M} \right)$ is easy to compute. However, the drawback is that it might suffer from non-ignborable estimation bias if the subsample size is relatively small. To address this issue,
\cite{zhu2022feature} developed two improved methods for bias reduction. The first method is a jackknife-based bias-correction method. To be specific, define a delete-one estimator as $R^{2}_{(b)i}(\mathcal{M})$, which is the $R^{2}_{(b)}(\mathcal{M})$ statistic computed without the $i$-th observation. 
This leads to the jackknife estimator for the bias as $\widehat{\Delta}_{(b)}=n^{-1}(n-1) \sum_{i} R^{2}_{(b)}(\mathcal{M})-(n-1) R^{2}_{(b)i}(\mathcal{M})$. Then the jackknife-based bias-correction metric is defined as
$
R_{\mathrm{JBC}}^{2}\left(\mathcal{M}\right)=B^{-1} \sum_{k=1}^{B}\{R^{2}_{(b)}(\mathcal{M}
)-\widehat{\Delta}_{(b)}\}.
$
The second method is an aggregated moment method.  
This method first decomposes $R^{2}_{(b)}\left(\mathcal{M}\right)$ into several moment components, and then aggregates the components separately. To be more precise, note that $R$-Squared statistic is composed of three components, namely $\hat{\Sigma}_{\mathbb{X}(b)}^{\mathcal{M}}=n^{-1}(\mathbb{X}_{(b)}^{\mathcal{M}})^\top \mathbb{X}_{(b)}^{\mathcal{M}}, \hat{\Sigma}_{\mathbb{X} \mathbb{Y}(b)}^{\mathcal{M}}=n^{-1}(\mathbb{X}_{(b)}^{\mathcal{M}})^\top \mathbb{Y}_{(b)}$, and $\hat{\sigma}_{y(b)}^{2}=n^{-1}\|\mathbb{Y}_{(b)}-\overline{\mathbb{Y}}_{(b)}\|^{2}$. Averaging over all subsamples leads to 
$\hat{\Sigma}_{\mathbb{X}}^{\mathcal{M}}=B^{-1} \sum_{b=1}^{B} \hat{\Sigma}_{\mathbb{X}(k)}^{\mathcal{M}}, \hat{\Sigma}_{\mathbb{X Y}}^{\mathcal{M}}=B^{-1} \sum_{k=1}^{B} \hat{\Sigma}_{\mathbb{X Y}(k)}^{\mathcal{M}}$ and $\hat{\sigma}_{\mathbb{Y}}^{2}=B^{-1} \sum_{b=1}^{B} \hat{\sigma}_{y(b)}^{2}$. Then, an aggregated moment estimator is defined as
$
R_{\mathrm{AM}}^{2}\left(\mathcal{M}\right)={\hat{\sigma}_{\mathbb{Y}}^{-2}}(\hat{\Sigma}_{\mathbb{X Y}}^{\mathcal{M}})^\top(\hat{\Sigma}_{\mathbb{X}}^{\mathcal{M}}
)^{-1}(\hat{\Sigma}_{\mathbb{X Y}}^{\mathcal{M}}).
$
Both two methods can reduce the bias of $R^2_{\mathrm{OS}}(\mathcal{M})$ significantly without inflecting the asymptotic variance.

\csection{MINIBATCH RELATED TECHNIQUES}

\csubsection{A Selective Review on Statistical Optimization}

In statistical research, many estimation problems can ultimately be transformed into optimization problems. For example, for the generalized linear models (GLMs), one usually estimates the model parameters by maximizing the likelihood function \citep{nelder1972generalized}. The likelihood function is generally a sufficiently smooth and strongly convex function. Consequently, the Newton's method or Fisher's score method can be easily implemented.
Often the numerical convergence can be achieved in a few iterations \textcolor{black}{\citep{Shao2003}}.  
Therefore, researchers usually do not concern much about the specific optimization process but directly study the statistical properties of the optimizer \citep{van2000asymptotic}.

However, with the rapid development of information technology, not only datasets are becoming increasingly large, but models are also becoming even more complex \citep{fan2020statistical}. 
As mentioned before, this poses two challenges to traditional optimization methods. 
First, the dimension of the model parameters $p$ can be very high. This would make it difficult to invert the $p \times  p$ Hessian matrix for Newton's type methods. 
Second, the dataset may be too large to be read into computer memory as a whole. 
Then the optimization methods based on whole datasets become no longer feasible.
For the former challenge, one can consider gradient-based first-order optimization methods, such as gradient descent (GD) method, quasi-Newton method, and conjugate gradient method \citep{beck2017first}.
For the latter challenge, one can load the data into the memory in a minibatch-wise manner, and then implement the algorithms based on these small minibatches. This leads to various minibatch-based methods. In particular, when the minibatches are generated randomly, they are also referred to as stochastic optimization methods \citep{First-order2020}, such as stochastic gradient descent (SGD). 
Due to the scalability of these minibatch-based first-order optimization methods, they are now widely used in large-scale learning tasks such as deep learning \citep{bottou2018optimization, simonyan2014very,he2016deep}. 

In addition to the popularity in practice, the theoretical properties of minibatch related algorithms have also attracted increasing attention from researchers. 
The early literature on solving optimization problems using the idea of stochastic approximation can be traced back to \cite{robbins1951stochastic} and \cite{kiefer1952stochastic}. 
In order to improve the efficiency of approximation, \cite{polyak1992acceleration} further proposed to perform averaging over the iterates, also known as PJR-averaging operation. 
More recently, \cite{moulines2011non} and \cite{bach2013non} investigated the SGD algorithm for objective functions with and without strong convexity, establishing the non-asymptotic convergence upper bounds.
For objective functions of finite sums, several variance reduction techniques have been found useful in achieving a faster convergence rate compared to the classical SGD \citep{roux2012stochastic,johnson2013accelerating,defazio2014saga}. 
To further accelerate the convergence rate for ill-conditioned problems, momentum based methods have also attracted great attention \citep{gitman2019understanding, assran2020convergence,liu2020improved}. 
Another line of research considered how to generate minibatches to improve the minibatch based GD \citep{needell2017batched,gower2019sgd,mishchenko2020random,gurbuzbalaban2021random}. 
Apart from research on minibatch related methods from an optimization perspective, there are also many studies conducted from a statistical perspective, paying more attention to characterizing the statistical properties of the resulting estimators. These works include but are not limited to \cite{toulis2017asymptotic}, \cite{chen2020statistical}, \cite{luo2020renewable}, \cite{zhu2023online}, and \cite{tang2023acceleration}.

\csubsection{Minibatch Gradient Descent Algorithms}

For high dimensional data analysis, various stochastic minibatch gradient descent (SMGD) methods have received increasing attention in recent literature due to their outstanding performances and relatively easier theoretical properties \citep{duchi2011adaptive,kingma2014adam}. The SMGD algorithms can be mainly categorized into two groups according to the generation of minibatch data. The first group assumes the minibatches are independently generated from the given sample with replacement \citep{OptinML}. Then the noise introduced by minibatch data can be viewed as conditionally independent. The second category assumes that the noises introduced by minibatch data form a martingale difference sequence. This assumption is particularly true for streaming data analysis \citep{Mou2020OnLS,An2020Yu,Chen2022StationaryBO}. 
Taking the SMGD studied in \citet{chen2020statistical} as a concrete example, they assume that the gradient noise from different minibatch data forms a martingale difference. Following the previous work of \citet{polyak1992acceleration}, the asymptotic distribution of the averaged SMGD estimator can be established. To conduct statistical inference of the averaged SMGD estimator, different inference procedures are proposed for both fixed dimension case and diverged dimension case. In the fixed dimension case, \citet{chen2020statistical} proposed a novel batch-means covariance estimator which can avoid computing the inverse of the Hessian matrix as compared with the naive plug-in estimator. In the high dimension case, \citet{chen2020statistical} proposed an online debiased lasso procedure to construct the confidence interval of element-wise regression coefficient.

Different from the SMGD algorithm, another way to generate minibatch data is the random partition, which is arguably the most popularly used minibatch method in offline real practice, since it has been well implemented by many standard deep learning programs such as TensorFlow and PyTorch. Since the minibatches form a partition of the whole sample data, they are no longer independent or conditionally independent with each other. As a result, it does not match the model assumption in the past literature \citep{bottou2018optimization,Bridging2020,First-order2020,Mou2020OnLS,An2020Yu,Chen2022StationaryBO}, which calls for theoretical investigation. To fill this theoretical gap, \citet{qi2023statistical} studied the properties of fixed minibatch gradient descent (FMGD) algorithm and the resulting estimator. Let $\mS=\{1,2,\dots,N\}$ be the index set of the whole sample. Let $Y_i\in\mathbb{R}^1$ be the response of interest and $X_i=(X_{i1},X_{i2},\dots,X_{ip})^\top\in\mathbb{R}^p$ be the associated $p$-dimensional predictor. Define the loss function evaluated at sample $i$ as $\ell(X_i,Y_i;\theta)$, where $\theta\in\mathbb{R}^{q}$ denotes the parameter. Then the global loss function can be constructed as $\mL(\theta)=N^{-1}\sum^N_{i=1} \ell(X_i,Y_i;\theta)$. The global estimator can be defined as $\widehat\theta = \operatorname{\argmin}\mL(\theta)$. Denote $\{\mS^{(t,m)}\}^M_{m=1}$ as the minibatch index sets in the $t$-th epoch. Then one should have $\mS=\bigcup_m\mS^{(t,m)}$ and $\mS^{(t,m_1)}\bigcap \mS^{(t,m_2)} = \emptyset$ for any $t\geq 1$ and $m_1\neq m_2$. For convenience, assume $N$ and $M$ are particularly designed so that $n=N/M$ is an integer and all minibatches have the same sample size as $|\mS^{(t,m)}|=n$. Then the updating formula of MGD can be expressed as 
\begin{eqnarray}
\label{eq:mgd}
    \widehat\theta^{(t,1)}&=&\widehat\theta^{(t-1,M)}-\alpha\dot{\mL}^{(t,1)}\Big(\widehat\theta^{(t-1,M)}\Big),\\
\widehat\theta^{(t,m)}&=&\widehat\theta^{(t,m-1)}-\alpha\dot{\mL}^{(t,m)}\Big(\widehat\theta^{(t,m-1)}\Big) \mbox{ for } 2\leq m \leq M,\nonumber
\end{eqnarray}
where $\alpha>0$ is the learning rate, $\mL^{(t,m)}(\theta)=n^{-1}\sum_{i\in\mS^{(t,m)}}\ell(X_i,Y_i;\theta)$ is the loss function for the $m$-th minibatch in the $t$-th epoch, and $\dot{\mL}^{(t,m)}(\theta)$ is the first-order derivatives of $\mL^{(t,m)}(\theta)$ with respect to $\theta$. 

To study the algorithm, \citet{qi2023statistical} first consider FMGD algorithm under the linear regression model with a fixed sample partition. Then the above updating formulas \eqref{eq:mgd} naturally form a linear system. Under appropriate technical assumptions, \citet{qi2023statistical} show that the FMGD estimator converges linearly to the stable solution of the linear system as
\begin{equation}
\Big\|\widehat\theta^{(t,m)}-\widehat\theta^{(m)}\Big\|\leq \rho_{\alpha,M}^{t-1}\Big\|\widehat\theta^{(0,m)}-\widehat\theta^{(m)}\Big\|,
\nonumber
\end{equation}
where $\rho_{\alpha,M}\in (0,1)$ is a contraction factor depending on $\alpha$ and $M$. The asymptotic normality result is also established by \citet{qi2023statistical}. However, \citet{qi2023statistical} find that the FMGD estimator is biased for any constant learning $\alpha >0$. To further reduce the error upper bound, \citet{qi2023statistical} consider the diminishing learning rate scheduling. As long as $\sum^{\infty}_{t=1}\alpha_t = \infty$ and $\sum^{\infty}_{t=1}\alpha^2_t<\infty$, the FMGD estimator should converge to the OLS estimator as $t\to\infty$. \citet{qi2023statistical} then extend their theoretical investigations to random partition with shuffling and general loss function, and similar results are established. 

\csubsection{Minibatch Gradient Descent with Momentum}

Despite the practical usefulness of the MGD algorithm, it still can be extremely time-consuming for large-scale statistical analysis with high dimensional parameters, particularly in the research field of deep learning \citep{kingma2014adam, he2016deep, goodfellow2016deep, devlin2018bert}. To address this issue, various improved algorithms have been proposed. One direction is to investigate the accelerated gradient descent algorithm. As a first-order optimization method, gradient descent (GD) does not require the computation of the second derivative (i.e., the Hessian matrix) of the objective function. 
However, due to the neglect of the second-order information of the objective function, the numerical convergence rate of the standard GD algorithm is often much slower than that of second-order optimization algorithms such as Newton's method. This is particularly true when the objective function is severely ill-conditioned. 
To address this issue, \cite{polyak1964} proposed a  so-called ``heavy-ball" method. This method utilizes not only the gradient of the current step but also the information from the previous step (i.e., momentum). 
Specifically, it updates the estimates as 
\begin{align*}
	\wh\theta^{(t)} = \wh\theta^{(t-1)}  - \alpha \dot\mL( \wh\theta^{(t-1)}  ) + \gamma \Big(\wh\theta^{(t-1)} - \wh\theta^{(t-2)}\Big),
\end{align*}
where $\wh\theta^{(t)} $ is the $t$-th estimate, $\dot\mL(\theta)$ is the gradient, $ \alpha >0$ is the learning rate, and $\gamma>0 $ is the momentum parameter. 
Further theoretical analysis by \cite{polyak1964} showed that the convergence rate could be much improved compared to the standard GD algorithm by this modification.
This method is now commonly referred to as the gradient descent with momentum (GDM).

The success of the momentum idea has attracted considerable attention from both theoretical and practical perspectives \citep{sutskever2013importance, goodfellow2016deep,bottou2018optimization}.
For example, \cite{nesterov1983method} proposed a method that uses the momentum and predicted gradient to update the parameters. 
By a similar idea, \cite{beck2009fast} developed an accelerated optimization algorithm for non-smooth objective functions. \cite{kingma2014adam} proposed an adaptive momentum method called ADAM, which is widely used in the fields of deep learning. \cite{cyrus2018robust} described a robust momentum method that generalizes the triple momentum method proposed in \cite{van2017fastest}. 
\cite{ma2018quasi} developed a more general variant called the quasi-hyperbolic momentum (QHM) algorithm, whose theoretical properties were further investigated by \cite{gitman2019understanding}.
In practice,  when the whole dataset is too large to be loaded into the memory, one has to process the data in a minibatch-by-minibatch manner. 
This is particularly true for many deep learning tasks \citep{krizhevsky2012imagenet,simonyan2014very,he2016deep}. This leads to various minibatch-based GDM (MGDM) methods.  
In fact, the MGDM methods have been incorporated into many important software libraries, including TensorFlow \citep{DBLP:journals/corr/AbadiABBCCCDDDG16} and PyTorch \citep{NEURIPS2019_bdbca288}.

In recent years, many efforts have been devoted to the theoretical analysis of the MGDM methods. 
Most of these studies investigated various MGDM methods from an optimization perspective, and have shown that the momentum term can effectively improve the numerical convergence \citep{gitman2019understanding, loizou2020momentum,liu2020improved, assran2020convergence}.
A recent work of \cite{tang2023acceleration} analyzed a PJR-averaging version of the MGDM method for general statistical optimization problems and established the asymptotic distribution of the resulting estimator.
It is worth noting that most of the theoretical studies mentioned above typically require that the minibatches are sampled independently and identically from the whole dataset (or the population distribution). 
However, in practice, such as in TensorFlow or PyTorch, minibatches are often obtained through random partition. 
As mentioned before, random partition means that the whole dataset is randomly partitioned into several non-overlapping minibatches. Unfortunately, minibatches generated in this way no longer satisfy the aforementioned requirements. 
To bridge the gap between theory and practice, \cite{gao2023asymptotic} considered the random partition based MGDM algorithm as a linear dynamical system:
\begin{align*}
	\wh\theta^{(t,m)} = \wh\theta^{(t,m-1)}  - \alpha\dot\mL_{(m)}(	\wh\theta^{(t,m-1)} ) + \gamma \Big(	\wh\theta^{(t,m-1)} -	\wh\theta^{(t,m-2)} \Big), \ \textup{for } 1\le m\le M,
\end{align*}
where $\wh\theta^{(t,m)} $ is the estimate from the $m$-th minibatch in the $t$-th epoch, and $\mL_{(m)}(\theta) $ is the gradient computed on the $m$-th minibatch $(1\le m\le M)$.

Based on the linear regression model, a closed form of the stable solution to the above linear dynamical system can be obtained. Specifically, let $\wh\theta^{(m)} $ be the stable solution corresponding to the $m$-th minibatch. 
Under appropriate conditions, one can obtain the following linear convergence result:
\begin{align*}
	\Big\|\widehat{\theta}^{(t, m)} - \widehat{\theta}^{(m)} \Big\| \le \Big(\rho_{\alpha, \gamma}^M + \epsilon_{n,t}\Big)^t \Big(\Big\|\widehat{\theta}^{(0, m)} - \widehat{\theta}^{(m)} \Big\| + \Big\|\widehat{\theta}^{(0, m-1)} - \widehat{\theta}^{(m-1)} \Big\|\Big),
\end{align*}
for each $ 1\le m\le M $, where  $\rho_{\alpha, \gamma}\in (0,1)$ is the contraction factor controlling the convergence rate, and $\epsilon_{n,t}$ is some small number. By choosing appropriate tuning parameters $\alpha$ and $\gamma$, one can achieve the minimal (and thus the optimal) contraction factor $\rho_{\min} = (\sqrt{\kappa}-1) / (\sqrt{\kappa} + 1)$, where $\kappa$ is the condition number of the least squares problem. 
In addition, \cite{gao2023asymptotic} established the asymptotic normality for the stable solution. The results showed that the stable solution can be statistically as efficient as the whole sample OLS estimator, as long as learning rate $\alpha$ is sufficiently small.
However, this interesting result relies on the least squares loss function. Investigating the MGDM algorithm based on a randomly partition strategy under general loss functions is a problem worth further exploration.

\csubsection{Communication Reduction for Minibatch Gradient Descent Algorithm}
 
Another research direction focuses on how to reduce communication costs during the training process of MGD algorithms. Massive datasets are usually stored on hard drives due to a limited storage capacity of the computation devices. However, the data need to be transmitted into the system RAM and/or graphical memory before gradient computation. Therefore, the communication cost during the training process of the MGD algorithm might also lead to expensive time consumption \citep{2017IOcost,2018IOcost}. The communication mechanism of MGD algorithm also causes another problem when computing in a CPU-GPU system, that is, the GPU idling problem. This problem refers to the phenomenon where the GPU waits for the CPU to perform data communication before gradient computation during the training process \citep{zinkevich2010parallelized,2016ICML,2019echoing}. It could become even worse when the number of updates is extremely large. Then how to reduce the communication cost during the training of MGD algorithms for faster computation becomes a problem of great interest \citep{bauer2011cudadma,2018IOcost,zhu2019efficient, ofeidis2022overview}.

This problem can be solved by either advanced hardware technology or algorithmic innovation. There has been remarkable technological advancement in the first direction. For example, the GPUDirect Storage technology developed by NVIDIA enables direct communication between the hard drive and the GPU graphical memory. Furthermore, their Remote Direct
Memory Access (RDMA) technology provides a direct path among different GPUs within a distributed system. As for the second direction, various methods have been proposed to either reduce the idling time or improve the communication efficiency. For example, \citet{2019echoing} proposed a data echoing method, which repeatedly computes the gradient on the currently loaded minibatch data before the next minibatch to be prepared. \citet{2020BA} further considered applying different data augmentation method to the currently loaded minibatch data. To reduce the idling time of GPU and  enhance computation efficiency, the idea of pre-loading and buffering has also received a lot of attention; see for example the data processing pipelines \citep{nitzberg1997collective,ma2003improving,bauer2011cudadma} and asynchronous data loading \citep{zhu2019efficient, ofeidis2022overview}. Despite their usefulness, the existing methods still suffer from several challenges. First, novel hardware technologies often require substantial engineering labors and financial expenses. As a result, these methods have not been extensively adopted by general practitioners and researchers for now. Second, since solving the GPU idling problem is quite engineering-oriented, most existing algorithms and training strategies lack theoretical analysis. Consequently, it is of great importance to investigate the theoretical properties of those methods under certain framework.

To address this issue, \citet{qi2023minibatch} consider a buffered minibatch gradient descent (BMGD) algorithm. The proposed BMGD algorithm consists of two steps, that is the buffering step and the computation step. In the buffering step, a large amount of data are loaded into the CPU memory. To this end, assume that the entire sample can be divided into $K$ non-overlapping blocks, which are referred to as the buffered data. Their indices are collected by $\mS_{r,k}$ ($1\leq r\leq R, 1\leq k\leq K$). For all $1\leq r\leq R$, one can obtain $\mS=\bigcup_k \mS_{r,k}$ and $\mS_{r,k_1}\cap \mS_{r, k_2}= \emptyset$ for $k_1\neq k_2$. Then in the computation step, a standard MGD updating procedure is applied on the buffered data. Assume that each buffered data can be further decomposed into $M$ minibatches. Let $\mS_{r,k}^{(t,m)}$ be the index set of the sample calculated on the $m$-th minibatch in the $t$-th epoch for the $k$-th buffer in the $r$-th iteration and assume that $|\mS_{r,k}^{(t,m)}| = n$ for all $r,k,t,m$. By inheriting the model assumptions in Section 4.1, the updating formula of the BMGD algorithm can be written as 
\begin{eqnarray}
    \label{eq:1}
\widehat{\theta}_{r,k}^{(t,1)}&=&\widehat{\theta}_{r,k}^{(t-1,M)}-\alpha\dot{\mL}_{r,k}^{(t,1)}\left(\widehat{\theta}_{r,k}^{(t-1,M)}\right),\nonumber\\
\widehat{\theta}_{r,k}^{(t,m)}&=&\widehat{\theta}_{r,k}^{(t,m-1)}-\alpha\dot{\mL}_{r,k}^{(t,m)}\left(\widehat{\theta}_{r,k}^{(t,m-1)}\right) \mbox{\ \ for\ \ } 2\leq m \leq M,
\end{eqnarray}
where $\alpha>0$ is the learning rate, and $\dot\mL_{r,k}^{(t,m)}(\theta)=n^{-1}\sum_{i\in\mS^{(t,m)}_{r,k}}\dot\ell(X_i,Y_i;\theta)$ is the gradient computed on minibatch $\mS^{(t,m)}_{r,k}$. The buffering idea of the BMGD algorithm can help reduce the GPU idling time and improve communication efficiency. A rigorous asymptotic theory was developed by \citet{qi2023minibatch} to support the BMGD method. Its applicability is also extended to the Polyak-Lojasiewicz (PL) function class, which not only contains a wide range of statistical models (e.g., the generalized linear model) but also some non-convex loss functions.

\csubsection{Learning Rate Scheduling}

One crucial component to the minibatch gradient descent algorithm and its variants is an appropriate learning rate scheduling. Despite previous efforts showing that the algorithm should converge under certain conditions, determining the correct learning rate in real practice largely relies on subjective judgment \citep{Sebastian2016overview}. If the learning rate is set inappropriately, it can lead to training failure or slow convergence. To address this issue, various scheduling approaches have been proposed to achieve adaptive adjustment of the learning rate. Examples include rule-based scheduling, such as the step decay scheduler \citep{NIPS2019_9635} and the ``reduce learning rate on plateau'' method \citep{nakamura2021learning}. These methods automatically reduce the learning rate when optimization encounters bottlenecks. Besides, \citet{duchi2011adaptive} proposed AdaGrad, which iteratively decreases the step size based on a pre-specified function. However, AdaGrad requires setting an additional parameter related to the learning rate, which needs to be subjectively chosen. Extensions of AdaGrad, such as RMSProp \textcolor{black}{\citep{mukkamala2017variants}} and AdaDelta \citep{zeiler2012adadelta}, have been proposed. RMSProp introduces a decay factor to adjust the weights of previous sample gradients. Moreover, Adam \citep{kingma2014adam} combines RMSProp with a momentum-based method called adaptive moment estimation. In Adam, both the step size and update direction are adjusted during each iteration. However, because the step sizes are adjusted without considering the loss function, the loss reduction obtained for each update step is suboptimal. Therefore, the convergence rate can still be further improved. New techniques have been introduced to adjust step sizes in gradient-based optimization methods. For instance, \citet{baydin2017online} introduced a ``learning rate for the learning rate'' hyperparameter, which is updated using gradient descent to adjust the learning rate. 
\citet{shu2022metalrschedulenet} have built an additional network to predict learning rate values in different iterations. However, those works bring more hyperparameters and thus result in more parameter tuning as well as more uncertainties.

From the perspective of optimization, the key to improving training algorithms lies in how to appropriately exploit 1st-order information (i.e., the gradient) and the 2nd-order information (i.e., the Hessian matrix). When the scale of parameters to be estimated is relatively small, the most common approach has been to apply the Newton's method.
For training neural networks with a large number of parameters, several generalized optimization methods \citep{NIPS2016_6286,2017Second,bergou2020subsampling,gargiani2020promise} have been proposed, inspired by the Newton--Raphson iteration.
Due to the high computational cost, existing studies have tried to approximate the Hessian matrix, including the Barzilai-Borwein method \citep{NIPS2016_6286}, subsampling methods \citep{2017Second, bergou2020subsampling}, and  generalized Gauss-Newton method \citep{gargiani2020promise}. However, most of those methods still involve the computation and storage of 1st- and 2nd-order derivatives. Thus, they may be less efficient or even not feasible practically, when the dimension of parameter estimation is extremely high \citep{sutskever2013training}.

In order to achieve automatic and nearly optimal optimization, \citet{yingqiu2021} propose a novel optimization method based on local quadratic approximation (LQA). Viewing the learning rate as a time-varying parameter, they treat the loss reduction as a function of the temporal learning rate. Then, the learning rate is dynamically adjusted through the maximization of the loss reduction. Their aim is to make use of 2nd-order derivative information to accelerate the optimization while avoiding calculating the 2nd-order derivatives directly. To this end, they combine techniques of Taylor expansion and quadratic approximation to propose an improved optimization algorithm with low computational costs. 

Denote the time-varying learning rate as $\alpha_{t,k}$, where $t$ and $k$ are indices of iteration and minibatch, respectively.
Given a loss function $\mL(X;\theta)$ with the model parameter $\theta$ and input sample $X$, let $\hat{\theta}^{(t,k)}$ and $\Delta \mL(\alpha_{t,k})$ denote the estimate and the loss reduction at the $k$-th minibatch of the $t$-th iteration of the training, respectively.
For simplicity, let $g_{t,k} = |\mS_k|^{-1} \sum_{i \in \mS_k} \dot\mL( \hat{\theta}^{(t,k)})$ denote the current gradient, where $\mS_k$ is the index set of the $k$-th minibatch.
Then, the LQA method adopts Taylor expansion to explore the loss reduction. From the Taylor expansion of $\mL (\theta)$ around $\hat{\theta}^{(t,k)}$, one can obtain the following approximation,
\begin{align} 
    \Delta \mL(\alpha_{t,k}) &= \frac{1}{|\mS_k|} \sum_{i\in \mS_k} \left\{\mL \left(X_i;\hat{\theta}^{(t,k)} -\alpha_{t,k} g_{t,k}\right) - \mL \left(X_i;\hat{\theta}^{(t,k)}\right) \right\} \nonumber\\
    &= - a_{t,k} \alpha_{t,k} + b_{t,k} \alpha_{t,k} + 
o ( \alpha^2_{t,k} g_{t,k}^\top g_{t,k} ),\label{Taylor_expansion}
\end{align}
where the two constants are given by $ a_{t,k} = |\mS_k|^{-1} \sum_{i \in \mS_k} \dot\mL \left(X_i;\hat{\theta}^{(t,k)}\right)g_{t,k} $ and $b_{t,k} = (2|\mS_k|)^{-1} \sum_{i \in \mS_k}  g_{t,k}^\top \ddot\mL \left(X_i;\hat{\theta}^{(t,k)}\right)g_{t,k}$
 with $\dot\mL( \hat{\theta}^{(t,k)})$ and $\ddot\mL ( \hat{\theta}^{(t,k)} )$ denoting the 1st- and 2nd-order derivatives of the local loss function, respectively. Since the higher order terms here are negligible, equation \eqref{Taylor_expansion} can be simply written as
$    \Delta \mL (\alpha_{t,k}) \approx -a_{t,k} \delta_{t,k} + b_{t,k} \delta^2_{t,k}$.
To optimize $\Delta \mL(\alpha_{t,k})$ with respect to $\alpha_{t,k}$, one can take the corresponding derivative of the loss reduction, which leads to the approximated optimal learning rate $\alpha^*_{t,k} = (2b_{t,k})^{-1}a_{t,k}$.
This suggests that once the coefficients $a_{t,k}$ and $b_{t,k}$ are estimated, a nearly optimal choice for the learning rate can thus be determined. In this way, the reduction in the loss function is nearly optimal for each batch step. As a consequence, the total number of iterations required for convergence can be significantly reduced, allowing for the algorithm to converge much faster than usual.

\csection{CONCLUDING REMARKS}

This paper gives a selective review about three categories of statistical computation methods for massive data analysis. 
Firstly, we considered the distributed computing methods, which provide a feasible solution when the data size is too large to be comfortably accommodated by one single computer. 
Secondly, we considered the subsampling methods, which are practically useful, when limited computing resources are available for a massive data analysis task.
Finally, we considered the minibatch gradient techniques, which have been widely used for the training of complicated models with a large number of parameters, such as deep neural network models.

To conclude this paper, we discuss here some potential future research directions. Firstly, we remark that all the methods reviewed in this work are developed for independent data.
On the other side, datasets with sophisticated dependence structure (e.g., spatial-temporal data) are commonly encountered in real practice. 
 Then how to conduct distributed computation for data with sophisticated dependent structure is an important research direction. Second, all the methods reviewed in this work are all developed for models with relatively limited dimension and simple structure. 
 On the other side, models with extremely high dimension and extremely complicated structure are increasingly available. This is particularly true for various well-celebrated deep learning models. 
 Then how to develop statistical computing theory for models with extremely high dimension and extremely sophisticated structure is another important research direction. 
 Third, all the methods reviewed in this work assume a global model  universally for all the data points with a common set of model parameters. However, this seems an obviously unrealistic assumption for dataset with a massive size.
Then how to conduct effective statistical learning with more flexible model parameters is also an important research direction.
 Lastly, we remark that the three categories of methods reviewed in this paper can be combined together for even better performance. See for example \cite{yu2022optimal}  and \cite{chen2022first}. 
 This is the last research direction worth pursuing.

\section*{Disclosure Statement}

No potential conflict of interest is reported by the authors.

\renewcommand{\refname}{\centering REFERENCES}
		
\bibliographystyle{apalike}
\bibliography{references2.bib}

\end{document}